\documentclass[12pt]{article}
\usepackage[utf8]{inputenc}
\usepackage{natbib}
\usepackage{amssymb}
\setcitestyle{authoryear, open={(},close={)}}

\usepackage{graphicx,caption,subcaption}
\usepackage{amsmath}
\usepackage{enumitem}
\usepackage{soul}
\usepackage{authblk}

\usepackage[margin=1.25in]{geometry}

\usepackage{caption}
\usepackage{xcolor}

\title{Wildland Fire Mid-story: A generative modeling approach for representative fuels
\thanks{Los Alamos National Laboratory LA-UR-23-25924}
}
\author{Grant Hutchings}
\author{James Gattiker}
\author{Braden Scherting}
\affil{Statistical Sciences, Los Alamos National Laboratory}
\date{June 28, 2023}

\begin{document}
\maketitle
\begin{abstract}
Computational models for understanding and predicting fire in wildland and managed lands are increasing in impact. Data characterizing the fuels and environment is needed to continue improvement in the fidelity and reliability of fire outcomes. This paper addresses a gap in the characterization and population of mid-story fuels, which are not easily observable either through traditional survey, where data collection is time consuming, or with remote sensing, where the mid-story is typically obscured by forest canopy. We present a methodology to address populating a mid-story using a generative model for fuel placement that captures key concepts of spatial density and heterogeneity that varies by regional or local environmental conditions. The advantage of using a parameterized generative model is the ability to calibrate (or `tune') the generated fuels based on comparison to limited observation datasets or with expert guidance, and we show how this generative model can balance information from these sources to capture the essential characteristics of the wildland fuels environment. In this paper we emphasize the connection of terrestrial LiDAR (TLS) as the observations used to calibrate of the generative model, as TLS is a promising method for supporting forest fuels assessment. Code for the methods in this paper is available. 
\\
\begin{center}\small{\textbf{Keywords:} Wildfire Modeling, Cox Process, Bayesian Model Calibration, Gaussian Process, Prescribed Fire, Environmental Assessment, Spatial Generative Model}\end{center}

\end{abstract}

%%%%%%%%%%%%%%%%%%%%%%%%%%%%%%%%%%%%%%%%%%%%%%%%%%%%%%%%%%%%%%%%%
\section{Introduction}

Wildland fire has historically been modeled as a physically-inspired and empirically calibrated spread rate model applied to bulk fuel assessments \citep{rothermel1972, andrews2018}. Computational models of wildland fire have been explored, and range from fire perimeter spread approximations to spatially resolved models of fuels, atmospheric dynamics, and physical phenomenology \citep{linn2002, linn2005, linn2020}. These computational simulation models have been employed extensively in the study of the behavior and impacts of wildfire, for example \cite{linn2012}. Modeling and simulation of prescribed fire introduces additional challenges. In this report, we focus on the need for detailed layout of fuels, which is a key aspect of achieving reliable predictions of fire outcome \citep{linn2020, linn2021, hiers2009}.

Plot-level summary has been the basis in the past for characterization of mid-story fuels, that is, the fuels under the main forest canopy. For specific prescribed fire planning, direct survey consistent with the protocols of the Forest Inventory and Analysis (FIA) program \citep{toney2007} are currently used in operational settings. The use of reference plots to impute broad-area coverage \citep{riley2016} has made estimated plot-level statistics available with broad coverage. Manual survey is time-consuming and expensive and summarizes plots of hundreds of square meters in a relatively small number of metrics, sufficient for some purposes of large-scale environmental monitoring \citep{tinkham2018applications}. 
However, the spatial resolution of technologies for analyzing the interactions between forest composition and wildland fire now exceeds the spatial resolution sought by typical existing survey protocols. Computational models such as FIRETEC accept input on the meter scale, currently, allowing for the specification of explicit tree positions, and spatial layout of heterogeneous mid-story fuels. There is growing cognisance of the quantitative connection between heterogeneity in fuels to fire outcomes \cite{knapp2006heterogeneity, linn2013modeling}, and the influence of small-scale (meter to tens of meters) heterogeneity in fuel structure on salient fire behavior \citep{parsons2017numerical, atchley2021effects}. Further, prescribed fire planning targets marginal conditions, where outcomes have been shown to be sensitive \citep{jonko2021}. Survey protocols leveraging remote sensing technologies are emerging to satisfy the need for spatial resolution in fuels for wildland fire. 

Light detection and ranging (LiDAR) has found a strong and growing role in environmental characterization. Areal LiDAR survey (ALS) achieves resolution sufficient to inventory individual trees (e.g., \cite{li2012new}). This, coupled with the large spatial extent of ALS deployed from aircraft, enables exact representation of overstory fuels for large swaths, although currently not for complete coverage over time. Recent work examines the ability of \emph{in-situ} terrestrial LiDAR scanning (TLS) to estimate plot-level summaries  \citep{pokswinski2021}, including tree identification and bulk characterization of mid-story \citep{anderson2021, silva2016, rowell2020, lp-vegpred23}. This report will describe results based on information extracted from TLS, but is agnostic to the method of mid-story observation, which could be based on manual survey, remote or \emph{in-situ} sensing imagery, multi-modal LiDAR, or other means. 

Trees can be automatically extracted from broad-scale remote sensing survey, and this abstraction of trees and canopy can be used to populate the corresponding fuels in simulation models. Recent work \citep{mcdanold2022} develops an approach to generating grass and litter surface fuels using a physically-motivated model that includes dependence on overstory and canopy. The mid-story is not typically characterized in a complete format that allows exact layout, although it is correlated with canopy and other spatial covariates. The perspective taken in this report is that mid-story fuels will be characterized from limited relevant observations of an ecosystem, and generated for the purposes of analysis and simulation models as \emph{representative}, rather than exact, fuel distribution.

This report addresses the interpretation and generation of heterogeneous mid-story fuels with a spatial statistical modeling approach. The primary goal is the definition and demonstration of a statistical model that can generate representative mid-story fuels. A spatial statistical model can be calibrated to observed data; that is, the settings of the model that make it most consistent with observations can be discovered, perhaps also informed by prior constraints on the settings that control the characteristics of the generated spatial patterns. There is a large family of spatial statistical models that might be used to define generative processes. Setting reasonable values for the parameters of any generative model is known to be a considerable challenge, especially as the model grows in complexity. This work takes the position of investigating a model that is complex enough to capture features of heterogeneous fuels distributions, while simple enough to infer parameters of the model proposed given observations, or a practical combination of expertise and observational constraints.

The next section will describe the family of methodology that will be applied for representing and generating heterogeneous fuels layout. The approach to model calibration is then presented, along with a description of the metrics that will be used in the calibration process. A review of our datasets and goals leads to a demonstration validating the ability to infer generative model settings using instances from the model to support an understanding and validation of the capability in principle, and finally the corresponding process of determining model settings from observed data and the application of the method to fuels generation is shown. We conclude by describing how this model can fit into a framework of fuel layouts including additional spatial covariates, specifically the dependence of mid-story to canopy and to urban features, i.e. roads.

%%%%%%%%%%%%%%%%%%%%%%%%%%%%%%%%%%%%%%%%%%%%%%%%%%%%%%%%%%%%%%%%%
\section{Point-Process Model}

The goal of this work is to demonstrate a model that can generate synthetic layouts of mid-story fuels that are representative of an environment's characteristics. By \emph{representative} fuels, we mean that the model output is similar to the target environment in established metrics that are chosen to inform characteristics of interest in fuels layouts, including structure and heterogeneity. We now describe some background of spatial models that lead to the capability of generating representative fuels. 

Spatial statistical models have a rich history in the description and analysis of environmental data \citep{banerjee2003hierarchical}. The methodology here comes from two strands of this literature: spatial point processes and binary models. 

\subsection{Spatial Point Processes}

There is a deep literature in point processes, this section will introduce the topic relevant to the application. Data that contains the locations of events of interest (in space and/or time) are known as a point pattern. The standard term \emph{event} corresponds here to the siting of a fuel element, which from different points of view could represent a simple spatial sample of fuel, or be interpreted as distinct features such as shrubs. Suppose we observe the locations of $n$ plants $\{s_i\}_{n}$ in some plot or domain $D\in \mathbb{R}^2$. The natural model for these data is a point process $\mathcal{P}$, and we write $\{s_i\}_{n}\sim \mathcal{P}$. A commonly used point process is the Poisson process, with spatial variability controlled by an \textit{intensity function} $\gamma(\mathbf{s})$. The intensity function determines the density of points throughout the domain. Specifically, the expected number of points in a region $B\subset D$ (i.e, ``random counting measure'') is $\int_{B}\gamma(\mathbf{s})d\mathbf{s}$. When the intensity is constant, $\gamma(\mathbf{s})=\gamma$, we obtain a homogeneous Poisson process (HPP); the more general nonhomogeneous Poisson process (NHPP) is controlled by a non-constant intensity function; see \cite{moller2007modern} for further explanation and examples.

\subsection{Binary Mosaic Models}
 
The mid-story fuel spatial pattern can be represented as a binary mosaic model (BMM). In this approach, the observational data and corresponding model output are a 2-dimensional binary present/absent pattern over the domain. Here, the BMM is defined as a point process laying out the centers of disks, with the disk radii following a distribution. The layout of centers is spatially heterogeneous according to an underlying intensity distribution.

Approximating the map of mid-story fuels by a binary continuum constructed from the union of overlapping compact sets, for example in Fig.~\ref{fig:gmm_results}, is known generally as a \textit{germ-grain model}: germs (centers) follow a stochastic point process, and grains are given by compact sets which emanate from the germs and may depend on the point process. The \textit{boolean model} is special case of the germ-grain model that arises when the germs follow a homogeneous Poisson process, parameterized by a constant intensity $\gamma$, and the compact sets are mutually independent \citep{molchanov1997statistics} forming the binary mosaic. In previous work, inference on this model will typically target three parameters: the first two moments of the radii distribution and the scalar intensity value, which is assumed to be constant across the domain. This model can be extended to incorporate spatial heterogeneity through the use of an NHPP on the germs, yielding a spatially-indexed intensity surface $\gamma(s)$. Such a surface may be a parametric or nonparametric function of spatial locations. Nonparametric NHPPs can be difficult to estimate, but they obviate the problematic task of positing a parametric model for intensity \textit{a priori}. See ch. 8 of \cite{banerjee2003hierarchical} for an overview.

\cite{pielou1964spatial} was perhaps the first to draw attention to the use of binary mosaics in ecology. However, it was \cite{diggle1981binary} who first proposed an explicit generative model for binary mosaics in vegetation coverage map settings. Using a now widely-studied binary mosaic map of heather shrubs, he proposed a minimum contrast estimation strategy for inferring the homogeneous Poisson intensity and three Weibull parameters controlling radii. This analysis succeeded in estimating the parameters and generating realizations from the fitted models.

Inference on the boolean model for binary mosaics is expanded by \cite{moller2010likelihood}. This simulation-based likelihood inference method relies on a pre-specified Poisson process; the authors consider only homogeneous intensities. Our approach extends both results by permitting the underlying process intensity to vary over the domain, thereby enabling the modeling of more complicated vegetation structures. The approach is naturally extended to allow intensities that vary as a function of observed covariates, that is, spatial maps that indicate higher or lower expectation for observing fuel. 

\subsection{Spatial Generative Model}

We next present the details of the non-homogeneous Boolean process used to model the binary map, which is in the Gaussian Cox process family. The model is presented in component-wise manner that reflects the generative process; deviations between the formal statistical model and generative/sampling procedures are noted where relevant. 

The specification of a union of disks used to approximate the binary map requires the specification of locations of disk centers and the radii of the respective disks. We first specify the model for the number of disks $n$ and their respective center locations $s_1,\dots,s_n$, henceforth "points". 
The total number of points is sampled 
\[n \sim Pois(\lambda*A_D),\] 
where $A_D$ is the area of the domain and $\lambda$ represents points per unit area. We assume the layout of points in the domain $D$ is controlled by a smooth and continuous relative intensity function $\omega(s):D \to \mathbb{R}^+$. This is achieved by modeling $\omega$ as an appropriately-transformed Gaussian Process (GP), 
\begin{equation}\label{eqn:ppintensity}
    W(s) \sim \mathcal{GP(\mathbf{0}, C_{\rho})}
\end{equation}
\begin{equation}
    \gamma(s) = f(W(s))
\end{equation}
where $C_\rho$ is a covariance matrix generated by an appropriate covariance function with hyperparameter $\rho$, $f$ is a function mapping from $\mathbb{R} \to \mathbb{R}^+$.
Taking $f(s)=\exp(s)$ corresponds to a Log-Gaussian Cox process.

The proposed generative process takes 
$f(s)=logistic(s)$, which allows the interpretation of the relative intensity function as a probability. We model the relative intensity function $\omega(s)$, rather than the intensity function $\gamma(s)$ for convenience in the generative procedure. For the relative intensity function, $E[n_B] \neq \int_B \omega(s)ds$, rather $\forall s \in D,\; P(s_i=s)=\omega(s)$. 

Given a value $\rho = \hat{\rho}$, a process realization $W$ may be sampled from the GP. A process realization $W$ is function-valued, but a discrete approximation is obtained by computing the covariance function on discrete grid $d\times d$, yielding covariance matrix $\Sigma_{\hat{\rho}}$, and sampling from the multivariate normal distribution $\mathcal{N}(\mathbf{0}, \Sigma_{\hat{\rho}} )$. Transforming this draw by $f$ gives the intensity function. Conditional on the relative intensity function, the joint density of the number of points and point locations is
\begin{align}
\label{eqn:poissonpdf}
    s_1,\dots, s_n, n &\sim \prod_{i}\omega(s_i)\times \frac{(\lambda A_D)^n\exp(-\omega A_D)}{n!}.
\end{align}
The points are collected in the set $\Psi = \{s_1,\dots,s_n\}$, which is a sample from the NHPP.

We assume the locations and sizes are independent, so disk radii may be sampled from a distribution on $\mathbb{R}^+$. It is arguable that a spatial dependence exists for disk radii, but we will show that the inference process including a single spatial field can be challenging. Adding complexity to this model introduces significant challenges in inference, and information at this level of detail is not typically available for prior specification of distribution. We choose the truncated Normal distribution bounded below by zero:
\begin{align}
    \label{eqn:normaldisks}
    r_1,\dots,r_n\sim N^+(\mu, \sigma^2) 
\end{align}
Conditional on $r_i$, disk $i$ is the set $\xi_i = \{s\in\mathbb{R}^2: ||s-s_i|| \leq r_i\}$ and the binary map is
\begin{align*}
    \Xi &= \bigcup_{i:s_i\in \Psi} \xi_i.
\end{align*}

%%%%%%%%%%%%%%%%%%%%%%%%%%%%%%%%%%%%%%%%%%%%%%%%%%%%%%%%%%%%%%%%%
\section{Generative Model Calibration}

This section describes operational details of calibrating the generative model, including: generating spatial realizations from a model with specified parameters, metrics used for summarizing and ultimately comparing a spatial pattern, and the discovery of model parameters using the Bayesian model calibration framework. 

Generative modeling implements a stochastic map from unobservable parameters and observable covariates to data. These models are useful for encoding an approximation to the true data generating mechanism, systematically characterizing uncertainty, and are able to simulate new data. They also enable specification and estimation of parameters that control general data features relevant to the research question. This is useful when we wish to generate synthetic data with particular properties, but wish to integrate over or ignore other characteristics such as specific location or orientation. Developing appropriate feature sets capturing, or ignoring, qualities of the data is an application driven development process.

\subsection{Generating Model Realizations}\label{sec:gen_model_real}
A stochastic outcome of the model, that is, a spatial layout, is referred to a \emph{realization} of the model conditional on the parameter settings. 
In the generative setting, we use the Gaussian Process model intensity function in Eq.~\ref{eqn:ppintensity}, referenced on a $d \times d$ grid $\mathcal{X}_D$, to predict the intensity function at randomly proposed test points. To produce a binary map, candidate points are proposed and accepted or rejected based on the predicted intensity function. Given a candidate location $\tilde{s}=(x,y)$ drawn uniformly on $D$, the intensity function value $W(\tilde{s})|\mathcal{X}_D,\mathcal{C}_{\rho}$ is predicted from the Gaussian Process model. The point $\tilde{s}$ is accepted with probability $p=\omega(\tilde{s})=f(W(\tilde{s})|\mathcal{X}_D,\mathcal{C}_{\rho})$. This process is repeated until $n$ candidate points are accepted, where $n$ is a Poisson random variable. Each point is the center of a disc of points in the BMM. The disc radius will be drawn from the truncated Normal distribution of Eq.~\ref{eqn:normaldisks}. The outcome is the BMM shown, for example, in Fig.~\ref{fig:priorpost}, where the top row shows realizations of different parameters drawn from their ranges. In practice, $d$ should be large enough such that the points in $\mathcal{X}_D$ are able to capture the heterogeneity in the field observations. 

\subsection{Descriptive metrics }

Observational data and model output can be quantified and compared through extracted metrics describing features of the data relevant to the application, in this case describing spatial qualities such as area, perimeter, and topological features such as counts of connected components and holes. We summarize the geometric geospatial object by a vector of relevant summaries of the binary mosaic, $y=\eta(\Xi)$.

The metrics used are described in Table~\ref{tab:metrics}, with some comments on procedural details. This is not intended to be a final or exhaustive list of useful features, but rather a sufficient set for demonstration. Extensions are possible; one potentially useful example is metrics related to transect segment summaries from manual survey. 
\begin{table}[]
    \centering
    \footnotesize
\begin{tabular}{|p{0.11\linewidth}|p{0.9\linewidth}|}
\hline
    area & From a binary image, observed or derived from observations, area is the proportion of occupied pixels in the domain. 
    \vspace{3pt}
    \newline For a generated representation composed of disks on the domain, we use a Monte Carlo estimate for the area of the union of the $n$ disks in the generated data; this provides the flexibility to compute local areas and statistics of spatial auto-correlation described below.  \\
\hline
    perimeter & From a binary image, observed or derived from observations, perimeter is the number of occupied pixels adjacent to unoccupied pixels, with a Pythagorean correction for boundary pixels that abut occupied pixels diagonally. 
    \vspace{3pt}
    \newline For a generated representation, the following approximation is used: Arrange points $n_i$ along the circumference of each disk; discard any circumference points that lie within another disk; calculate the approximate perimeter as the proportion of points retained times the analytic total disk perimeters. A circumference point lies within another disk if the distance between the point and the center of another disk, $s_j$, is less than $r_j$. 
    \\
\hline
    NCC & Number of distinct connected components is easily calculated from raster data.
    \vspace{3pt}
    \newline For a generated representation, we construct a graph object using the \emph{igraph} R package. Graph nodes are given by disk centers; edge $e_{ij}=1$ if $d(s_i, s_j)\leq r_i + r_j$ and $e_ij=0$ otherwise. Given a graph, the package provides efficient functionality for computing the number of connected components. 
    \\
\hline
    holes & From a binary image, the number of holes is calculated by performing the same NCC operation on unoccupied cells and subtracting the number of CCs on the boundary.
    \vspace{3pt}
    \newline To compute the number of holes given only disk locations and radii, we use tools from the \emph{TDA} (topological data analysis) R package.  \\
\hline
    grid areas & 
    Computing local areas on subsets of the domain enables us to investigate how area is distributed or correlated within the domain by computing metrics of these sub-domains. 
    \vspace{3pt}
    \newline For this work we use a regular grid of sub-domains of size $1m^2$.
        \begin{itemize}[noitemsep,topsep=0pt]
        \item Moran's I and Geary's C — measures of spatial auto-correlation based on adjacency; several versions of each statistic may be obtained by changing the size of the sub-area and by changing how adjacency is defined (e.g., Moore vs Von Neumann neighborhood, neighborhood range, weights, etc.). 
        \item sum of sub-domain areas — this is an approximation to total area
        \item number of full sub-domains
        \item number of empty sub-domains
        \item sample variance of sub-domain specific areas 
    \end{itemize} \\
\hline
    empirical parameter estimates & 
    Empirical estimates of some model parameters are available from disk-based data. 
    \vspace{3pt}
    \newline With the number of observed fuel elements $n$, and the vector of observed disk radii $\boldsymbol{r}$:
        \begin{itemize}[noitemsep,topsep=0pt]
        \item $\hat{\lambda} = n/(d_{X}*d_{Y})$. 
        \item $\hat{\mu}=mean(\boldsymbol{r})$
        \item $\hat{\sigma}=sd(\boldsymbol{r})$.
    \end{itemize}
    Incorporating these estimates as metrics encourages parameters consistent with those estimated from the data. \\
\hline
\end{tabular}
\caption{Metrics used to compare observed and generated binary fuels patterns}
\label{tab:metrics}
\end{table}

\subsection{Bayesian Model Calibration}

\begin{figure}
    \centering
    \includegraphics[width=0.5\textwidth]{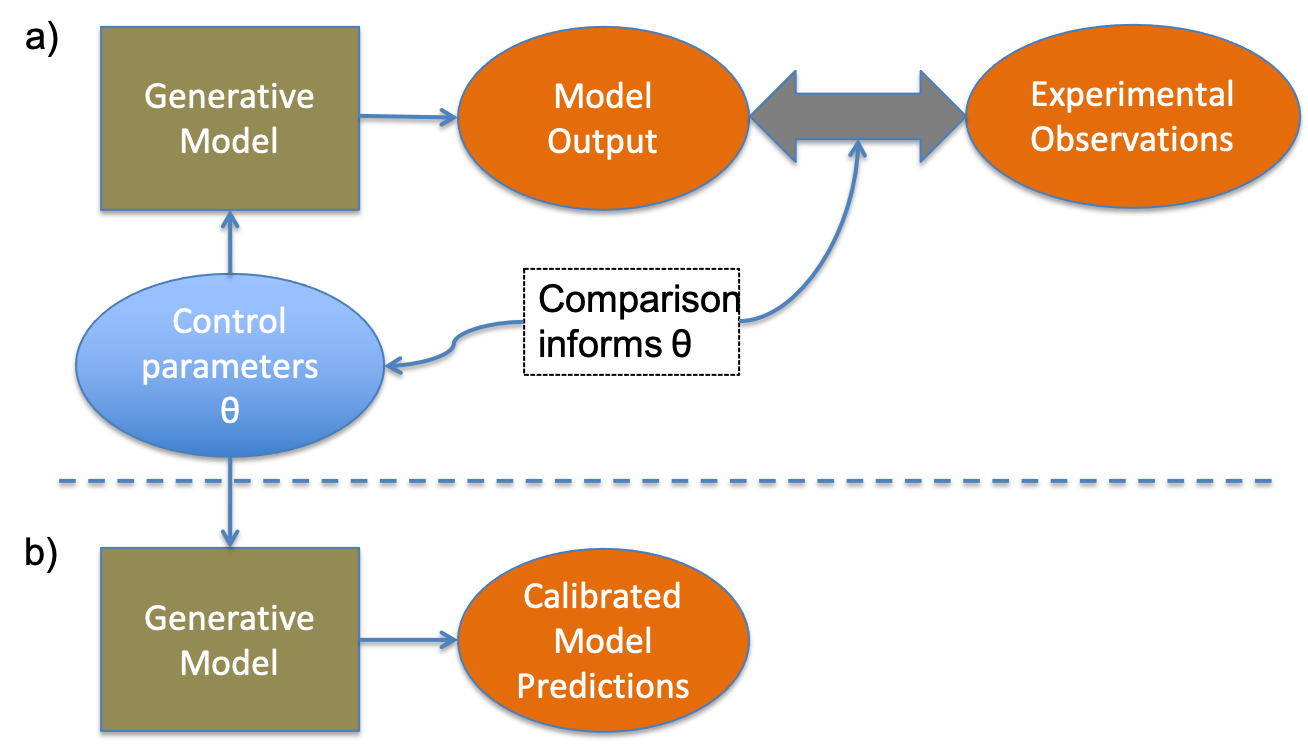}
    \caption{Model calibration concept. a) Simulations from a generative model across a prior range of control parameters $\theta$ are compared to observations, resulting in refined knowledge of $\theta$ distributions, and then b) the posterior distributions for $\theta$ lead to predictions with appropriate uncertainty / variability. }
    \label{fig:CalCartoon}
\end{figure}

The calibration process is represented graphically in Fig~\ref{fig:CalCartoon}.
Given a parametric generative model or forward simulation, model calibration refers to the process of estimating model parameter settings that give the most consistent output, or in other terms \emph{fitting} the model. Considering uncertainty, \emph{calibration} of the generative model to data is determining a posterior distribution of the model parameters given the observation data. New predictions from the calibrated model are made with the inferred parameter distributions for $\theta$. 

We cast this in a standard Bayesian calibration framework \citep{santner2018}. A posterior distribution with respect to the metrics $y=\eta(\Xi)$ is defined. The metric set $y^{obs}$ is extracted from observations of the natural environment. The corresponding set of metrics $y^{gen}$ can be extracted from a model generated example. These can then be compared through a likelihood function. Here, we assume a likelihood function associated with a multivariate normal distribution on the metrics, given the values of the parameters $\theta$:
\begin{equation}\label{eq:lh}
\L(y^{obs} | \theta) = (2\pi)^{k/2}det(\Sigma_{y|\theta})^{-1/2}e^{-\frac{1}{2} \sum_i\sum_j (y^{obs}_i - y^{gen}_j)^T\Sigma_{y|\theta}^{-1}(y^{obs}_i - y^{gen}_j)},
\end{equation}
where $k$ is the length of metrics vector. The covariance matrix $\Sigma_{y|\theta}$ is not known in practice and must be estimated, which is discussed in Section \ref{sec:results_model_calib}.
The next step in Bayesian calibration is to specify priors on the parameters $\theta$. In this implementation the default prior on lengthscale of the GP intensity function is Uniform, with bounds chosen based on the domain size. The prior on the disk mean is $N(1.5,0.5^2)$, truncated to $[0,3]$, and the prior on the variance is $\Gamma(1,.001)$ which has a peak near zero. These priors are reasonable defaults for many of the datasets we have worked with, but may need to be adjusted for different ecosystems. Together we denote the prior on the parameters as $\pi(\theta)$. The Bayesian expression of the posterior distribution for the parameters for a given set of data-derived metrics is:
\begin{equation} \label{eq:post}
P(\theta | y^{obs}) \propto L(y^{obs}|\theta) * \pi(\theta)
\end{equation}
The above expressions are un-normalized, presented as such since the normalization constants are not necessary operationally for inference. In principle, Eq.~\ref{eq:post} can be optimized to find the maximum \emph{a posteriori} value of the parameters, or it can be calibrated with Markov chain Monte-Carlo (MCMC) to sample the posterior distribution (we will not rehearse the MCMC algorithm here, it is easily available in references such as \cite{gelmanBDA}, which also includes substantial advice on practice). 

%%%%%%%%%%%%%%%%%%%%%%%%%%%%%%%%%%%%%%%%%%%%%%%%%%%%%%%%%%%%%%%%%
\section{Model Calibration Process and Results}\label{sec:results_model_calib}

An initial step in verifying the soundness of the procedure proposed is to infer the model based on so-called \emph{perfect data} - data generated from the model, but treated as observations. Through that process the capability can be assessed quantitatively as well as qualitatively, albeit for the ideal case of data known to be consistent with model assumptions. We then proceed to outcomes of the same model inference procedure applied to real-world observations. There several aspects to the statistical calibration process with some procedural complexity, some of which are unique to this application. This section will give some insight into practical considerations for calibrating this generative model to data, and how prior domain information can be introduced to aid inference. 

\subsection{Model Inference Process with Generated Data}

\begin{figure}
    \centering
    \includegraphics[width=0.75\textwidth]{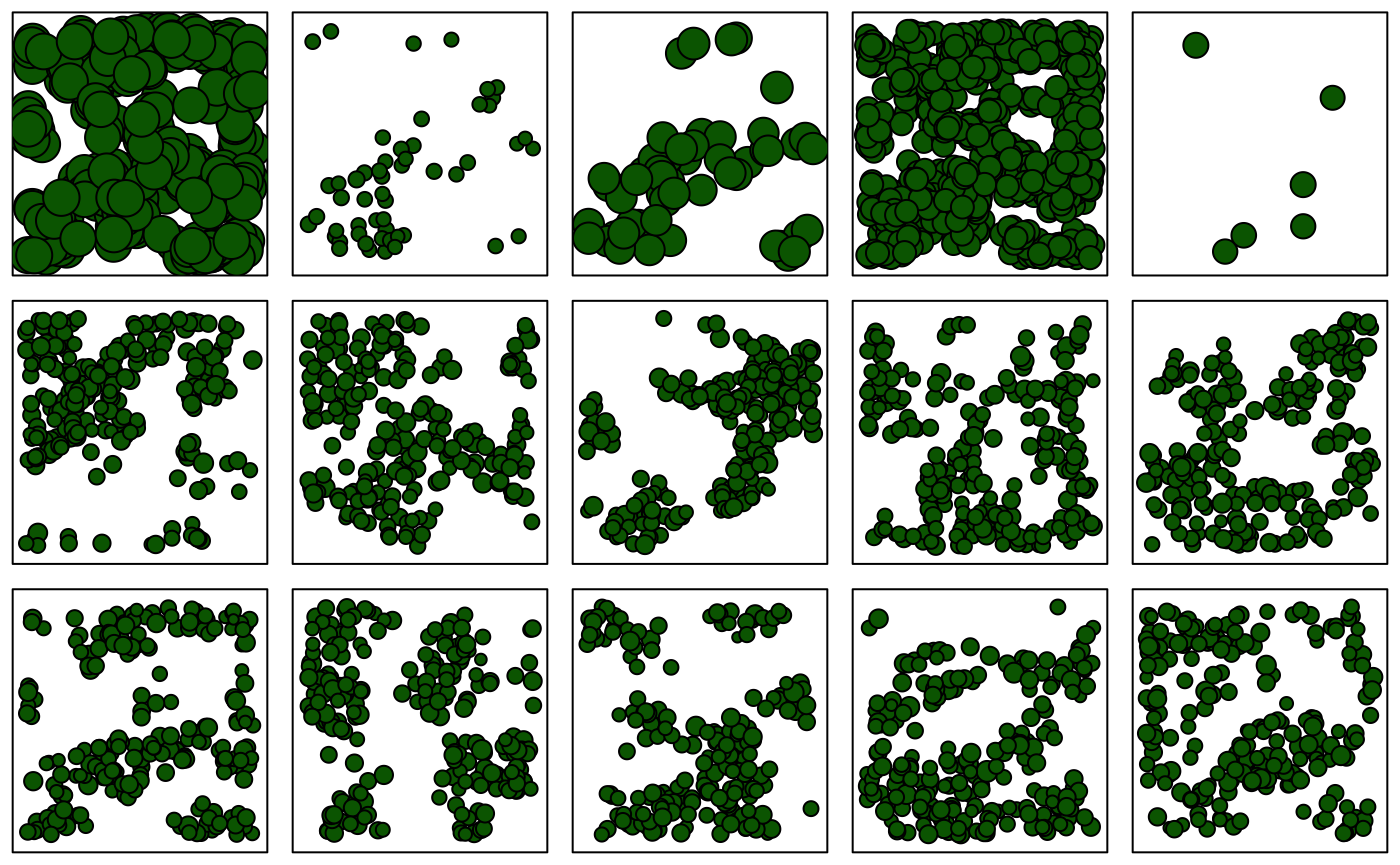}
    \caption{Impact of parameters on generative binary layouts. Top row: five parameter sets from their prior ranges demonstrates breadth of generative behavior; middle: synthetically generated layouts from one setting of parameters; bottom row: the constrained model behavior associated with the calibrated model, showing 5 generated outcomes from parameter sets drawn from the calibrated posterior distribution. }
    \label{fig:priorpost}
\end{figure}

We will first show the outcome of the calibration process, then fill in additional underlying technical and procedural details. 
Recall that this generative model has four parameters: the length scale of the Gaussian process intensity function $\rho$, the point density per square meter $\lambda$, and the mean $\mu$ and standard deviation $\sigma$ of the Normal distribution for disk radius. Generated patterns using values of these parameters drawn from parameter prior distributions is shown in the panels on the top row of Fig.~\ref{fig:priorpost}. There is clear qualitative difference in the character of the fuels layouts generated with different parameter settings. To validate that the metric set used allows the identification of these settings through comparison with observation, the panels in the middle row of Fig.~\ref{fig:priorpost} show an example of model-generated data as synthetic observation that will be used in model calibration, where all five are generated using the same parameter set. The lower row panels show fuels generated from samples of calibrated parameter distributions, which are qualitatively similar to the training data, and we will show the quantitative connection in model parameters.

\subsubsection{Stochastic Likelihood Evaluation}

In this application, find settings of parameters $\theta=\{\rho,\lambda,\mu,\sigma\}$ is complicated by the model output being non-deterministic, where the generated spatial binary map is a stochastic outcome for a given a setting of parameters. As defined previously, we denote a generated binary map as $\Xi$, and the metrics generated from that binary map as $y=\eta(\Xi)$. The observation is $y^{obs}$ (as well as the model-generated pseudo observation where used), and a generated instance from the model is $y^{gen}_{\theta}$. 
For a deterministic model, MCMC samples the posterior distribution of parameters, summarized in Eq.~\ref{eq:post}, with the sole model outcome ($J=1$) $y_{gen}$. For a stochastic model, each realization has a different spatial outcome and corresponding metrics set, resulting in a sample from a corresponding stochastic likelihood. The likelihood in Eq.~\ref{eq:lh} acts to estimate an expectation of the likelihood over multiple realizations, although still resulting in a stochastic estimate. The need for multiple realizations from a Cox process for robust parameter estimation has been documented \citep{cox_realizations}. This model has the same need for multiple realizations, especially for estimating the intensity function length scale $\rho$.

There is a trade-off between the cost associated with the number of model realizations used, and the algorithmic performance associated with the noisy estimate of $E[L(y^{obs} | \theta)]$. For such estimates of the likelihood, algorithmic methods such as MCMC can encounter optimistic tail values through random variation, making it difficult efficiently sample the posterior. Our approach to addressing this issue with MCMC is to re-calculate the current value for $L(y^{obs} | \theta)$ for each proposed MCMC step. In practice, we have found that, over a range of test problems with different parameter settings, 25 realizations are sufficient to effectively constrain the posterior while keeping the computational cost as low as possible. Some parameter values may require more samples due to larger variability in $y^{gen}$; this will ultimately depend on the details of the data and domain. 

Preliminary studies in optimization have shown a greater intractability for optimizers to converge with stochastic model estimates as compared to MCMC. For this reason, optimization is not explored as an inference tool in this work.

\subsubsection{Covariance Estimation}\label{sec:cov_est}

\begin{figure}
    \centering
    \includegraphics[width=0.75\textwidth]{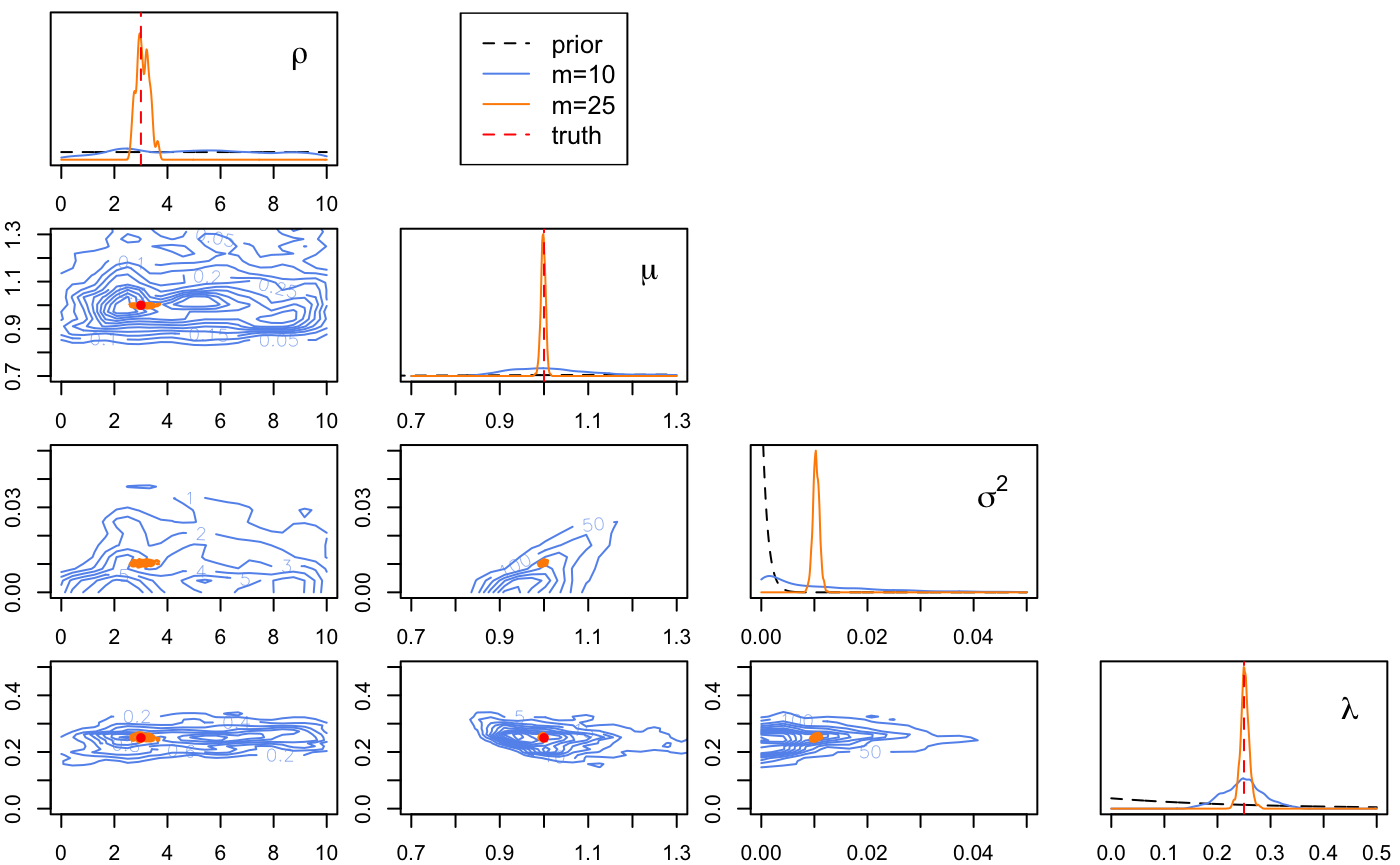}
    \caption{Model prior distributions (dotted) and calibrated posterior distributions for the parameters for the intensity function length scale, disk radius mean and variance, and overall intensity scaling average density. The parameters values used to generate the synthetic observations are shown in red vertical lines. Two posteriors are shown to demonstrate the value of observed data.}
    \label{fig:postdist}
\end{figure}

In Eq.~\ref{eq:lh} we defined a normal likelihood on the $m$ observed plots $y^{obs}_i\;i=1,...,m$, and explicitly denote that the covariance matrix is dependent on $\theta$ because of the stochastic nature of the data generating process for some $\theta$. As $\Sigma_{y| \theta}$ is unknown, an obvious plug-in estimate is the empirical covariance matrix of $y^{obs}_i;\;i=1,...,m$. This works well when $m$ is reasonably large. We have found that $m=25$ is sufficient for estimation of domains consistent with Fig.~\ref{fig:priorpost}, resulting in posterior uncertainty is significantly reduced from prior uncertainty for all parameters and posterior modes are consistent with their true settings. For smaller observed datasets of around $m\leq10$ we find that, all else equal, posterior uncertainty can be quite large and posterior modes can miss the true parameter settings. A representative example is shown in Fig.~\ref{fig:postdist}. Posterior distributions are much more sharply peaked with $m=25$ observations, and for all four parameters posterior modes are near the truth. For $m=10$ however, posterior uncertainty is significantly greater and inference for $\sigma$ and $\lambda$ may be biased towards smaller estimates.

There are two factors at play when comparing posterior estimates from the $m=10$ and $m=25$ case. First is that the sum likelihood has more information for larger $m$. Second is that the estimate of $\Sigma_{y| \theta}$ is improved with larger $m$. The former is not to be dismissed, however we have found that the latter may be the more significant effect. For our representative example in Figs.~(\ref{fig:priorpost}, \ref{fig:postdist}) we found that inference with $m=10$ using the covariance matrix estimated with $m=25$ is nearly identical to inference using $m=25$ for both estimation of the covariance and calibration. Improving covariance estimation can significantly improve inference without collecting more data.

To improve covariance estimation for these smaller data applications we have observed benefit in an approach of augmenting the observed dataset $y_{obs}$ with simulated data generated from random samples of $\theta$ from informed prior distributions. The augmented dataset is used only for covariance estimation, rather than likelihood calculations. Informed prior distributions can be developed naturally from the available observations, with empirical parameter estimates $\hat{\lambda} = n/(d_{X}*d_{Y}),\hat{\mu}=mean(r),\hat{\sigma}=sd(r)$.

To augment the observed data: draw $m^*$ samples $\boldsymbol{\lambda}_s \sim N(\hat{\lambda},.1\hat{\lambda})$, $\boldsymbol{\mu}_s \sim N(\hat{\mu},.1\hat{\mu})$, $\boldsymbol{\sigma}_s \sim N(\hat{\sigma},.1\hat{\sigma})$. The $10\%$ standard error is designed to account for the fact that our empirical estimates are subject to the random data $y^{obs}$. We cannot define a simple empirical estimate for $\rho$ so we sample $\boldsymbol{\rho}_s$ from weakly informative distribution such as a Uniform or Gamma with large variance. For each of these $m^*$ samples $\boldsymbol{ \theta}_s=[\boldsymbol{\lambda}_s|\boldsymbol{\mu}_s|\boldsymbol{\sigma}_s|\boldsymbol{\rho}_s]$ generate $K$ sets of random metrics $y_{gen}^k(\boldsymbol{ \theta}_m);\;k=1,...,K$. The covariance matrix is then estimated as the empirical covariance over these $m^*\times K$ sampled metrics together with the $m$ observations.

\begin{figure}
    \centering
    \includegraphics[width=1\textwidth]{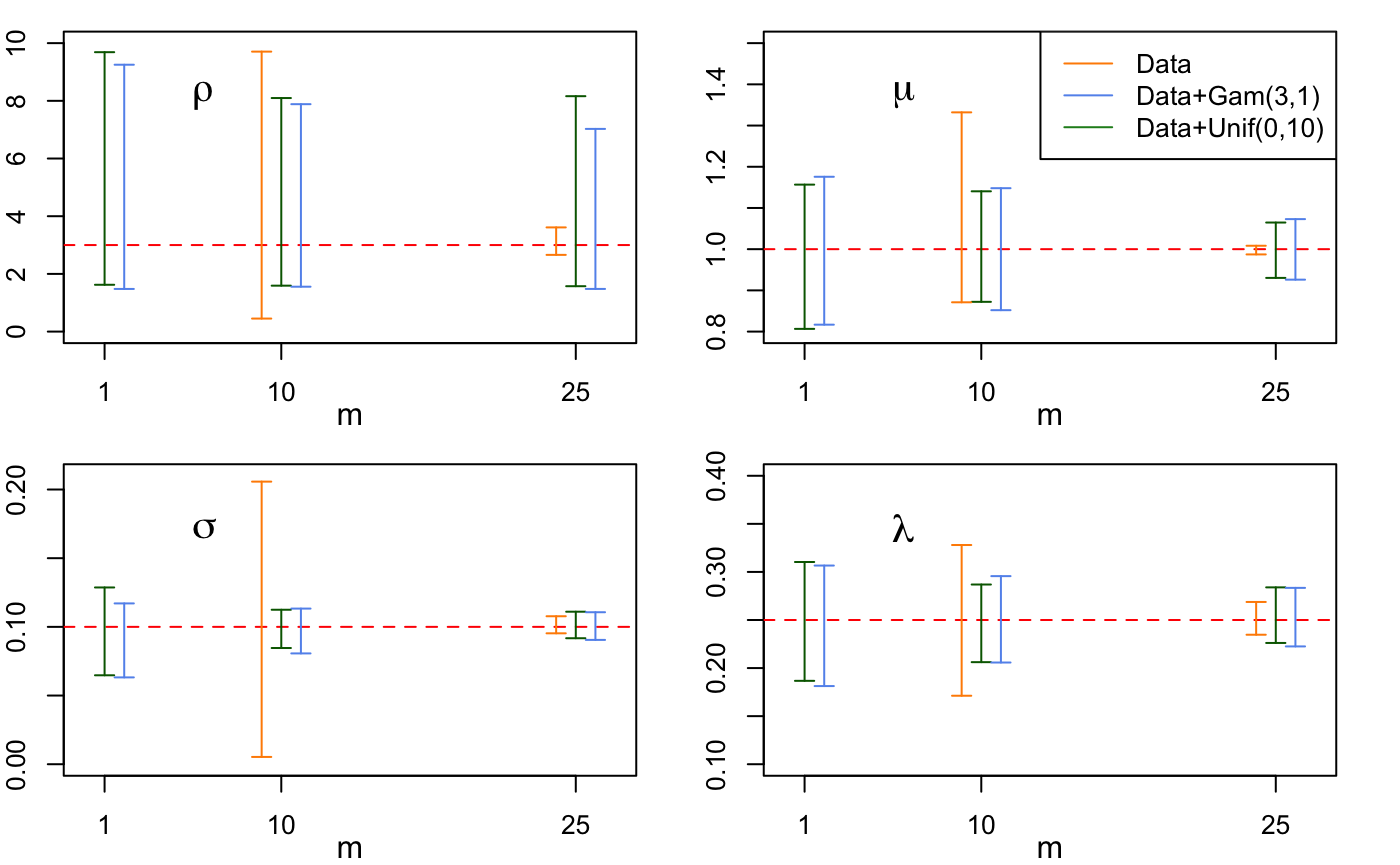}
    \caption{Posterior $95\%$ intervals over different observed dataset sizes and different priors for simulated data used to improve covariance estimation. Orange intervals are generated using only the data for covariance estimation. Green and Blue intervals use the observed data and simulated data to estimate the covariance. Augmenting observations with simulations for covariance estimation can significantly reduce posterior uncertainty when data is limited.}
    \label{fig:perfect_conf}
\end{figure}

Fig.~\ref{fig:perfect_conf} shows $95\%$ posterior confidence intervals over three values of $m \in \{1,10,25\}$ for different covariance matrix estimation approaches. In orange we have posterior intervals where only the $m$ available observations are used for covariance estimation. Blue and green intervals have incorporated simulated data in the estimation procedure. Both of these intervals use samples from the normal distributions defined above for $\lambda, \mu, \sigma$ but vary the prior for $\rho$. We find that sensitivity to the prior on $\rho$ is relatively small, with similar results from a Uniform distribution on $[0,10]$ and a more informative $\Gamma(3,1)$ distribution which has a mean at the true value. For this example, the conditional likelihood for $\rho$ with other parameters fixed at the truth is fairly flat.

This comparison demonstrates the connection between observation data availability and the ability to identify and constrain the parameter posterior distributions. It also shows that information regarding domain details, i.e. by a domain expert, can result in good outcomes even when relatively data limited.

\subsection{Calibration with Field Data}\label{sec:calib_obs}

Point patterns in spatial domains may arise either from direct observation or through post-processing of technical measurement. Direct observation is recording the locations of events of interest, often aided by GPS. When feasible, this form of survey produces high-fidelity measurements. However, it requires considerably more effort than either plot-level summaries or remotely-sensed surveys. A LiDAR scan produces a point cloud (i.e., a set of points) in 3D, but the points do not have the context of higher-level features such as plants, and the point clouds are processed to identify features of interest. For mid-story fuels, features of interest can be thought of as units of fuel organization, which, at different scales, could be bunch grass, bushes, or trees, or, more generally thought of as regions of high fuel density in the environment. The locations of the these features form a point pattern.

There are several possible pathways for extracting a heterogeneous spatial layout of fuels from observations of the environment. 
Monoculture ecosystems of chaparral or bunch grasses might be easily extracted from remotely sensed imagery or ALS as a coverage map that can be used directly as a binarized domain. A similar analysis is the extraction of the location and size of trees from imagery and ALS, for example \citep{treesfromALS}. However, the extraction of mid-story fuels is in general more difficult in an environment including an overstory. This section will show a simple method for extracting fuels elements from TLS scans, and used for calibration of the fuels model. 

\subsubsection{Terrestrial LiDAR Survey (TLS) }

\begin{figure}
    \centering
    \includegraphics[width=0.5\textwidth]{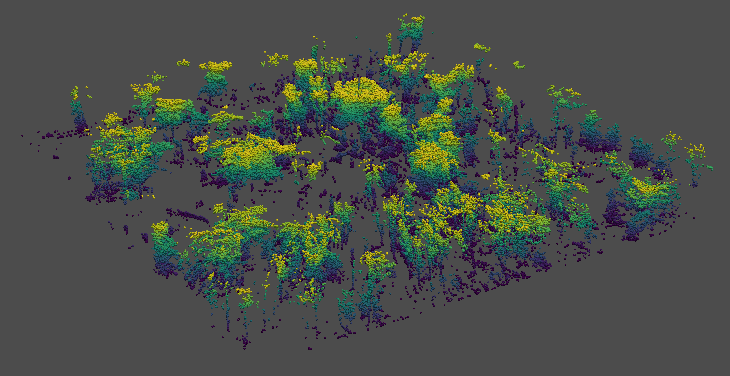}
    \caption{mid-story pointcloud used for example data extraction. This shows a ponderosa forest lidar scan, normalized for ground level, and segmented to exclude ground and canopy. }
    \label{fig:pointcloud}
\end{figure}

TLS generates a pointcloud of the scanned forest environment. The data collected here uses the Leica BLK360 instrument, deployed to sites in the Santa Fe National Forest. The demonstration site is ponderosa forest, which includes aspen at various maturity, gamble oak, and other typical mid-story flora. In this demonstration, the generated pointcloud is clipped to a 15m square domain centered at the scanner. In order to extract mid-story from the example scan, the ground is normalized to a ground plane, and the domain is segmented to the region from 0.1m to 3m. Fig.~\ref{fig:pointcloud} is a visualization of the resulting mid-story pointcloud data.
This report focuses on methodology detail rather than the environmental conditions or data collection methodology, and so we present this example dataset as representative data without defending its environmental properties.

\subsubsection{Data representation}
\begin{figure}
    \centering
    \includegraphics[width=0.5\textwidth]{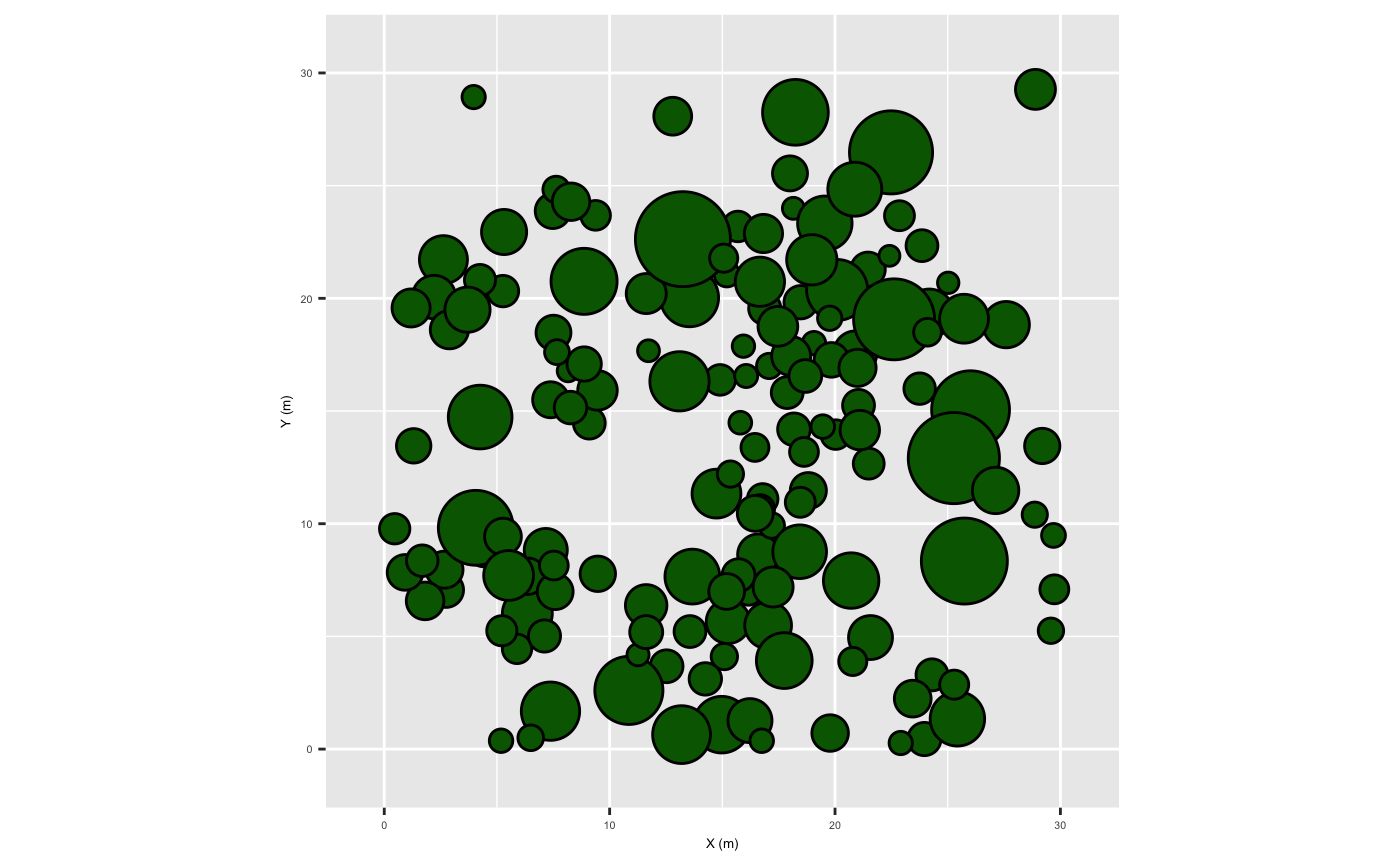}
    \caption{Pointcloud data projected down to the x-y plane, and the Gaussian mixture model fit to that data used to create a binary dataset. For presentation clarity, color distinguishes distributions and point membership. }
    \label{fig:gmm_results}
\end{figure}

We follow a simple reduction of the pointcloud data to spatially compact fuel units, which is the concept that motivated the spatial model representation. Since we are working in 2D for this demonstration, the z value of the data is disregarded, leaving the marginal projection to the ground plane. A Gaussian mixture model (GMM) of the 2D data is built using the python scikit library's BayesianGaussianMixture procedure, which fits a Dirichlet process model. The result a representation of the data as a number of Gaussian distribution, in this case constrained to be circular. A prior on the GMM fit controls the fuel data inferred to be in a reasonable range of radii. The inferred 2D fuels density is then discretized into a binary model of disks of 2 fitted GMM standard deviation radius centered at the corresponding GMM means. The fitted disks are shown in Fig.~\ref{fig:gmm_results}. Although this is an incomplete representation of the TLS mid-story information and there are alternative approaches for automating inventory, for example (\cite{lp-vegpred23}), and creating a characteristic representation, this description does reflect the variability of the major fuels elements in the domain, and in a format consistent with characterization of the mid-story as distinct units of fuel in the conceptual framework of the generative model presented. Specific of the algorithm will be highly dependent on the particulars of an application context of fuels generation; this background is a brief overview to introduce a path to generating the data shown in Fig.~\ref{fig:gmm_results}.

\subsubsection{Model Calibration and Fuels Generation}

\begin{figure}
    \centering
    \includegraphics[width=\textwidth]{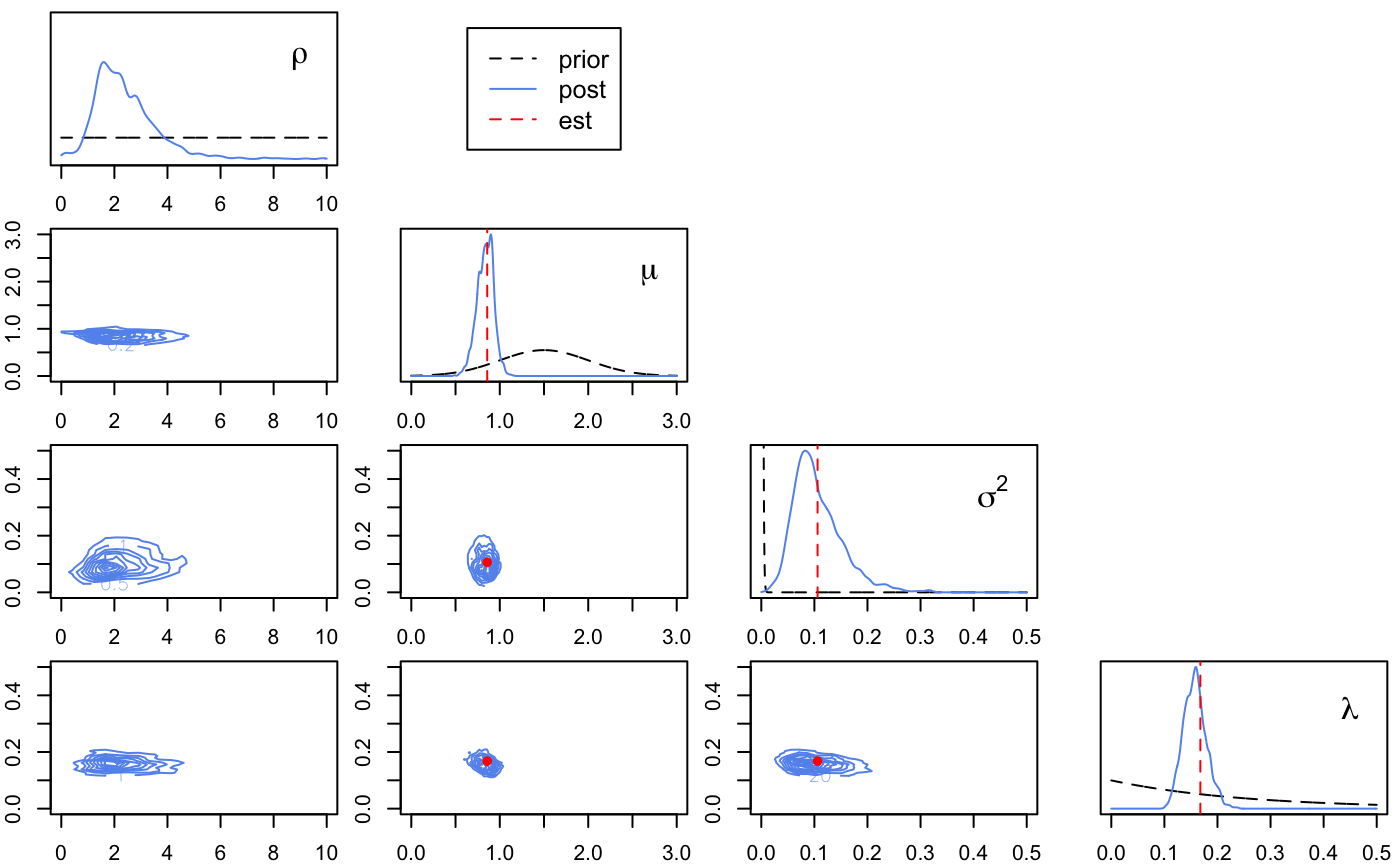}
    \caption{Parameter posterior distributions from model calibration against GMM representation of TLS mid-story. Empirical estimates from the single observation of $\mu,\sigma^2,\lambda$ are shown in red vertical lines.}
    \label{fig:gmm_calib}
\end{figure}

Calibrating the generative model with this disk-based BMM data corresponds directly to the calibration done with generated data. Calibration to the data in Fig.~\ref{fig:gmm_results} results in the posterior distributions shown in Fig.~\ref{fig:gmm_calib}. What we can see from these views of the 4-parameter joint distribution is that there is a significant degree of learning parameter posteriors from the presented example dataset, particularly for $\mu,\sigma,\lambda$. Covariance estimation for this calibration example was done using the data augmentation approach outlined in Section \ref{sec:cov_est} with $K=25$, $m^*=25$, and $\boldsymbol{\rho}_s$ drawn from a Uniform distribution on $[0,10]$.

\begin{figure}
    \centering
    \includegraphics[width=0.75\textwidth]{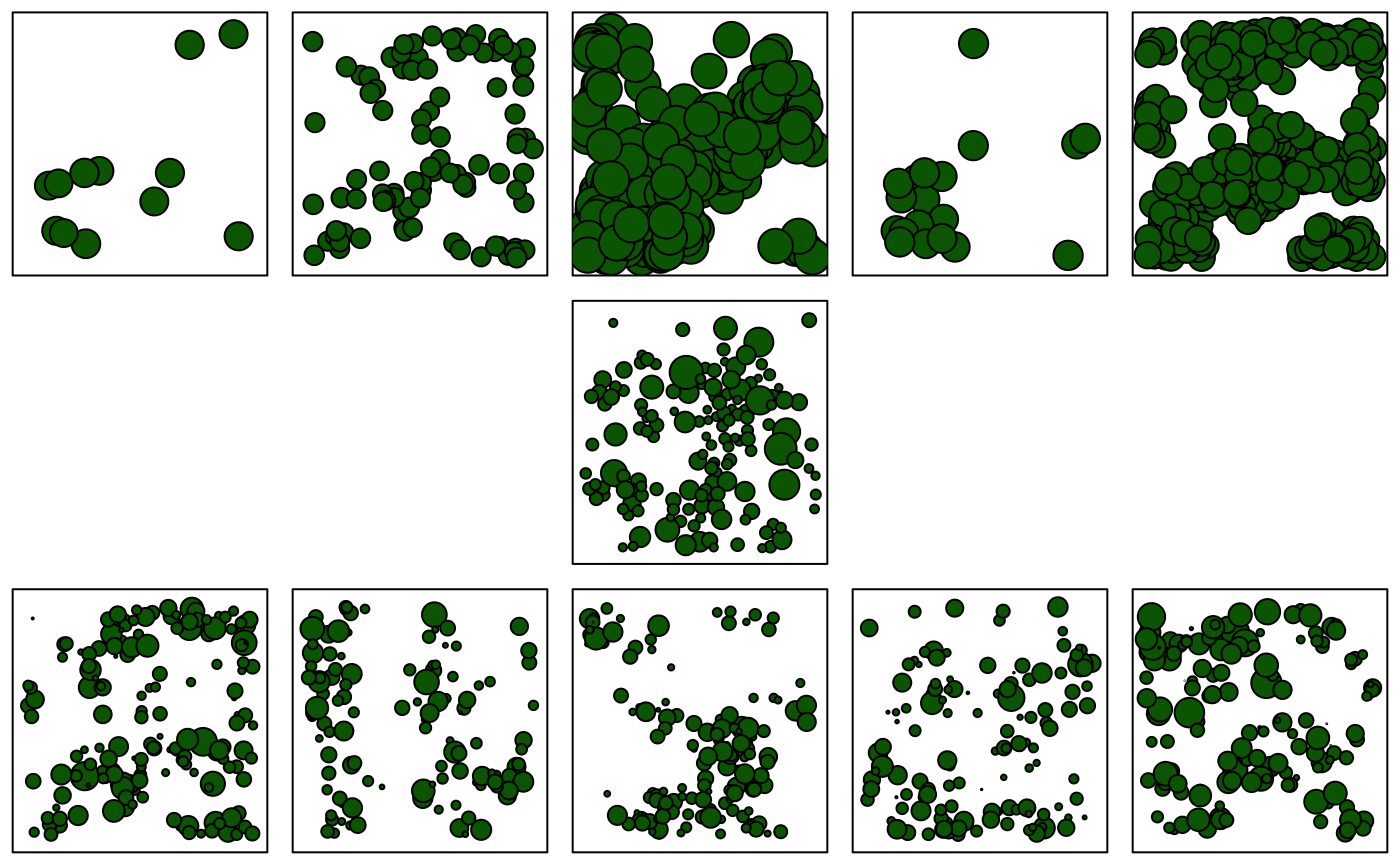}
    \caption{In the top row, five generated fuel layouts using parameters drawn from their prior distributions. The observed calibration data is shown in the middle row. In the bottom row, five generated fuel layouts from the calibrated model posterior distribution.}
    \label{fig:gmm_realiz}
\end{figure}

There is no ground truth in parameters in this case of data extracted from real TLS. Instead the evaluation is the qualitative performance of the fuels layout. This is shown in Fig.~\ref{fig:gmm_realiz} which, as before, shows some examples of the prior behavior of the model before calibration, the data used in calibration, and examples of the generative output from the calibrated model. Qualitative examination shows that the layouts of the calibrated model are reasonably representative, particularly in the goal of capturing the heterogeneity of the binarized maps. 

\section{A Framework for Multiple Spatial Influences}

\begin{figure}
     \centering
     \begin{subfigure}{0.475\textwidth} 
         \centering
         \includegraphics[width=.9\textwidth]{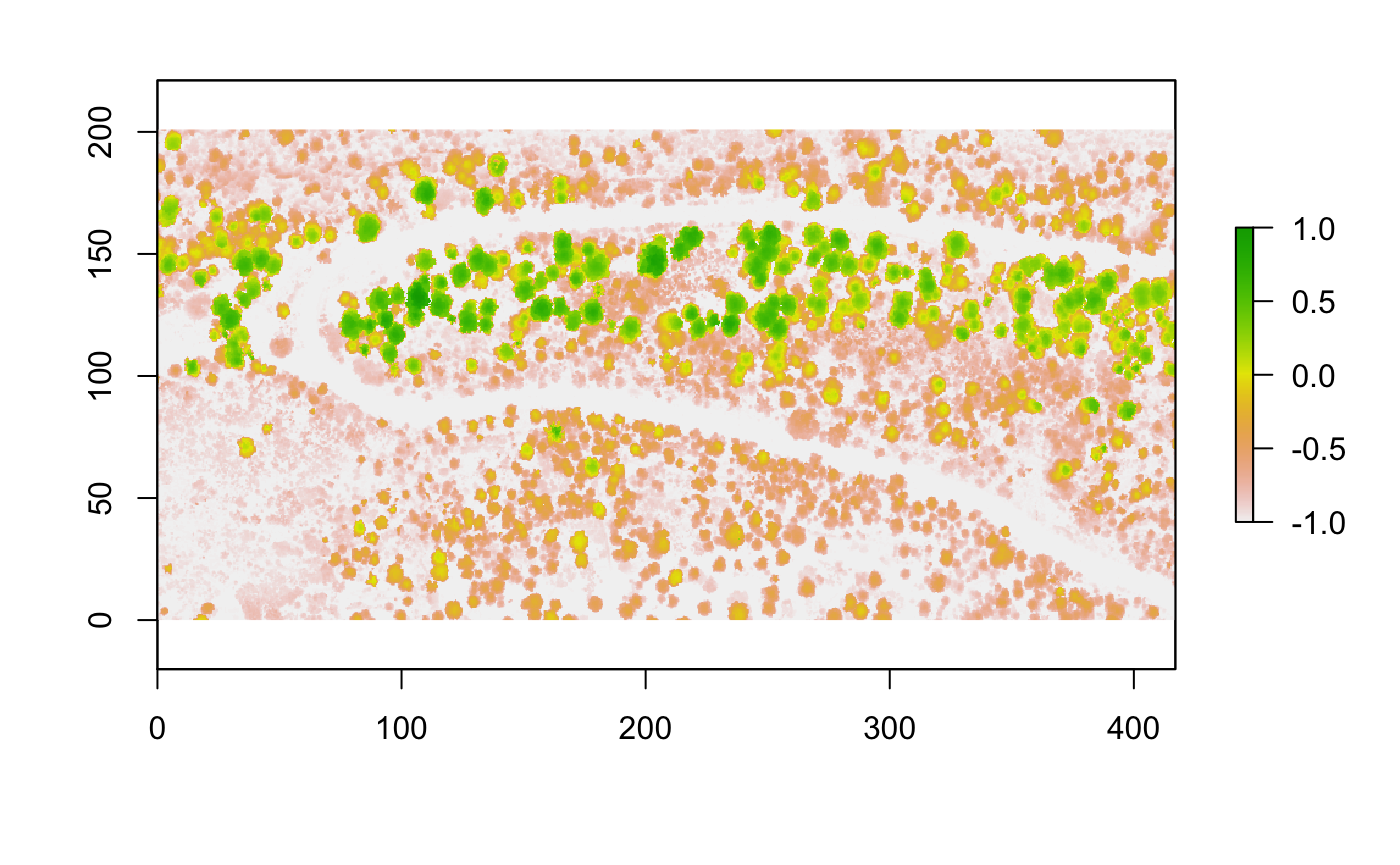}
         \caption{Canopy Height Model (CHM)}
         \label{fig:chm}
     \end{subfigure}
     \begin{subfigure}{0.475\textwidth} 
         \centering
         \includegraphics[width=.9\textwidth]{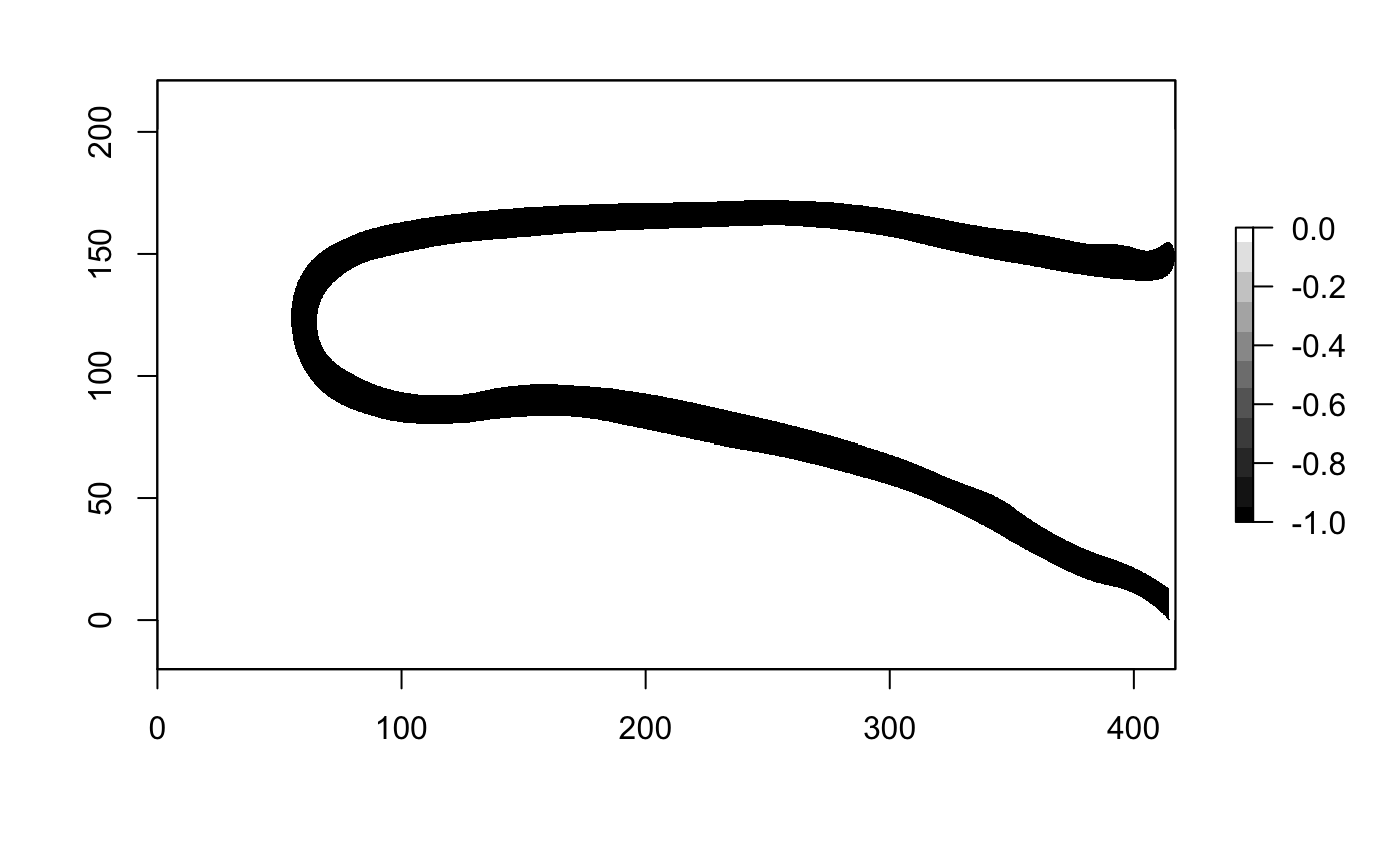}
         \caption{Location of road.}
         \label{fig:chm_road}
     \end{subfigure}
     \begin{subfigure}{0.475\textwidth} 
         \centering
         \includegraphics[width=.9\textwidth]{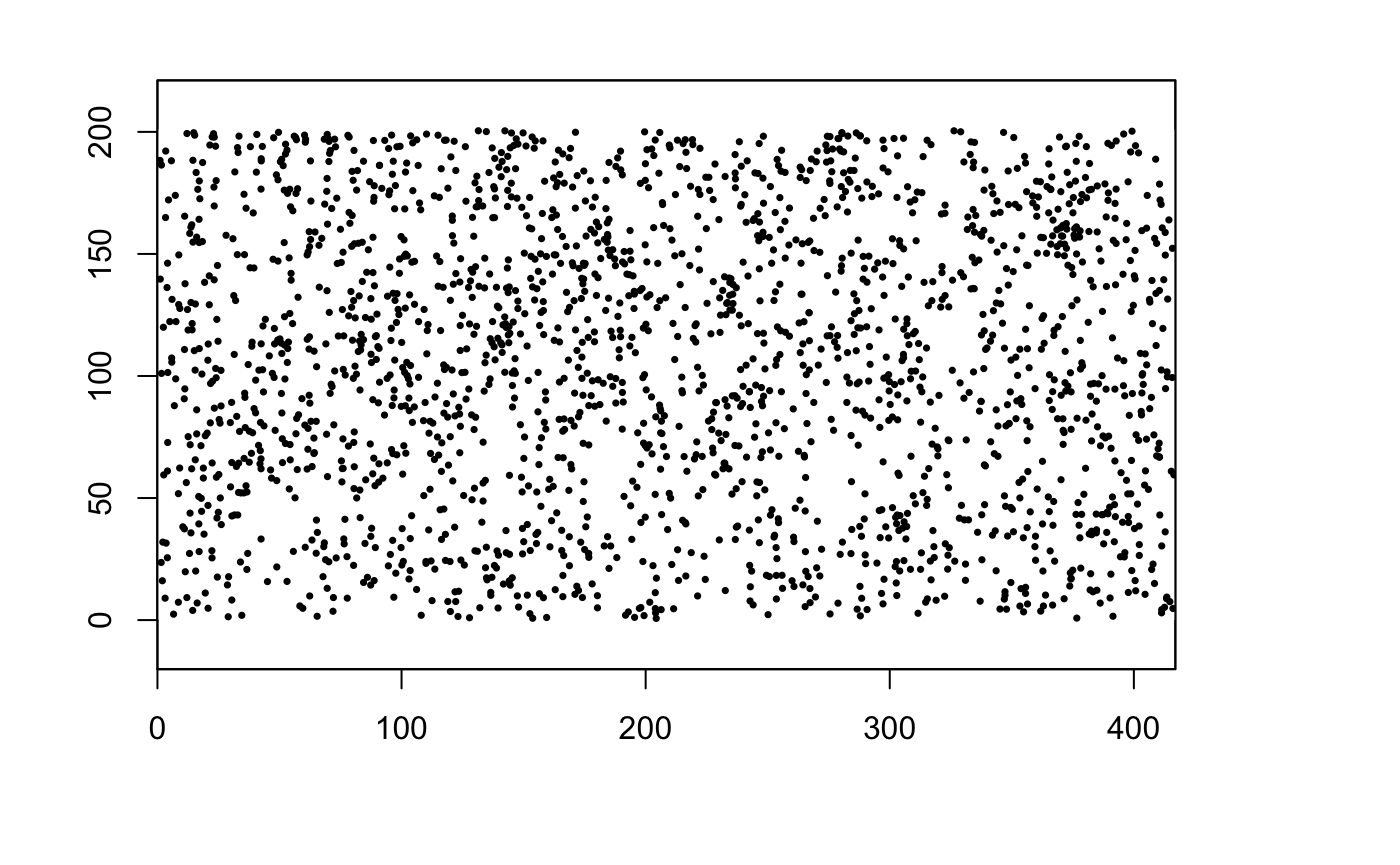}
         \caption{Generation without CHM influence.}
         \label{fig:chm_points_noB}
     \end{subfigure}
     \begin{subfigure}{0.475\textwidth}
         \centering
         \includegraphics[width=.9\textwidth]{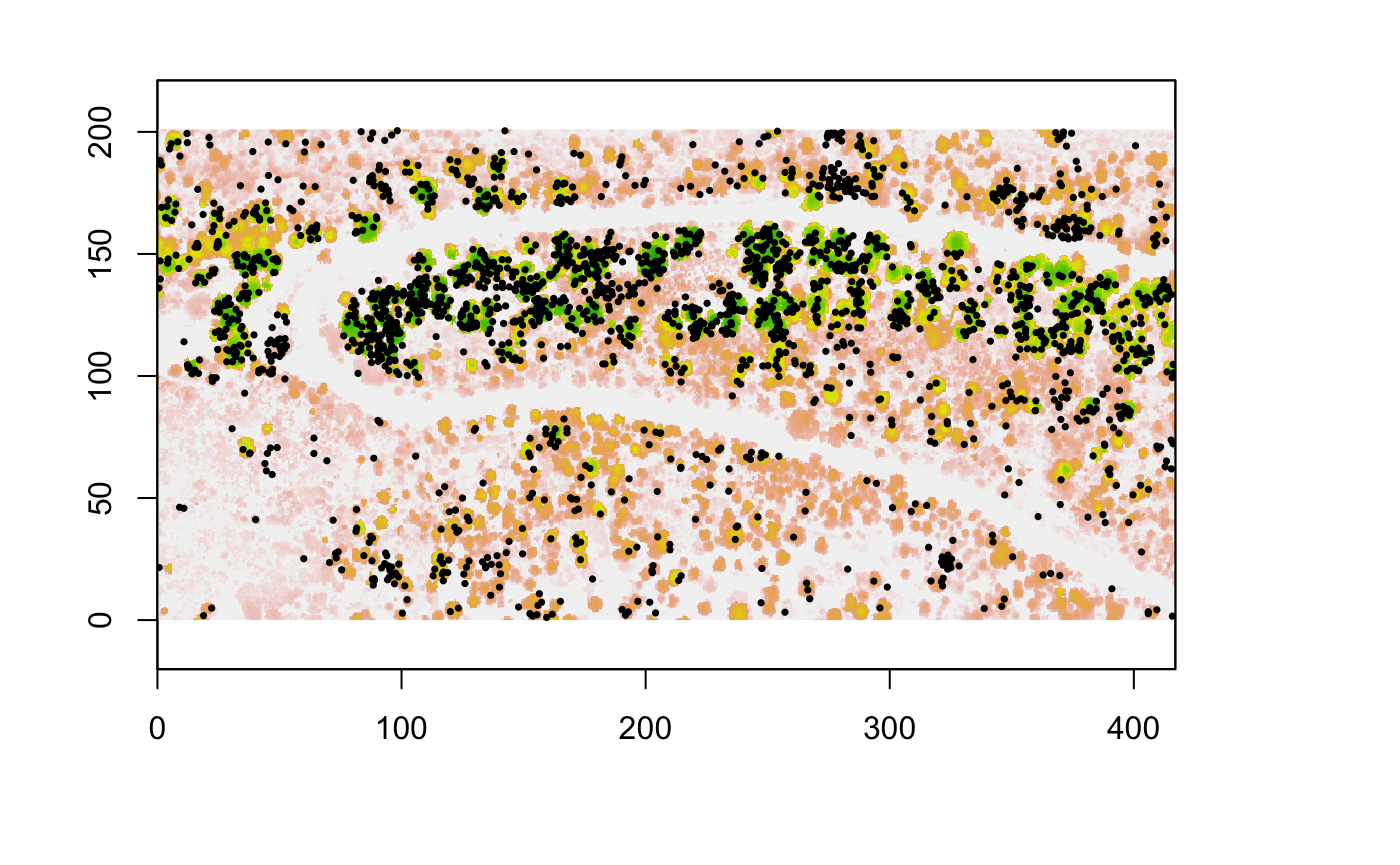}
         \caption{Generation with CHM influence.}
         \label{fig:chm_points_B}
     \end{subfigure}
        \caption{In panel (a) a canopy height model is shown, scaled to [-1,1]. Panel (b) shows the locations of the road, which should be set to zero probability of fuel generation. Panel (c) shows a realization from the data generating process given parameters $\tilde{\theta}$ with $\beta=(1,0,0)$ indicating no effect from the canopy. Panel (d) is a realization using the same $\tilde{\theta}$ but $\beta=(1,1,10)$ gives a strong correlation between canopy height and fuel placement. $\beta_2=10$ gives effectively zero probability of fuel placement at road locations indicated by $X_2$.}
        \label{fig:CHM_example}
\end{figure}

The model developed can be used along with other spatial covariates to provide a more contextually relevant layout. This can be accomplished by considering the generative model as only one component of a more general intensity function. Revisiting Eq.~\ref{eqn:ppintensity}, we can expand the intensity function heterogeneity component $W(s)$ to include other controlling terms, as:
\begin{equation}
    \omega(s) = f\big(\beta_0 W(s) + \sum_k \beta_k X(s)_k\big)
\end{equation}
That is, additional terms are added, where $X$ are spatial covariate fields, and $\beta$ are weights to control the relative impact of spatial field components. We will briefly show a generative example where $X_1$ is a canopy height model (CHM) over a domain, which we will take as an indicator of greater likelihood of finding a mid-story fuel element, i.e. a shrub. The canopy height model used for this example is shown in Fig. \ref{fig:chm}. In addition to a canopy height covariate, we also incorporate an $X_2$ of the location of a road, where fuel probability is zero. These are shown in Fig.~\ref{fig:chm_road}.

One important consideration in this model is the need for a convention for the semantics of the $X$ datasets. We will specify $X_k \in [-1,1]$, interpreted as proportional to a likelihood of fuel. $\beta_k$ is a positive multiplier interpreted as a relative scaling of the impact of indicator $k$. 
We will not investigate the ability to infer $\beta_k$ parameters, and instead will take them as supplied by a user guiding the system. To estimate $\beta$, data requirements would be substantially increased, and likely the need for expert-supplied guidance more critical, as well as complex. It is important to recognize that while the relative scaling of the $\beta_k$ controls the impact of indicators $X$ in the generated outcome, at the same time the overall sum of $\beta_k$ will impact the layout as the total is projected through the logistic function.

Fig.~\ref{fig:chm_points_noB} shows an example of placing fuel element centers in the domain of the CHM with the heterogeneity model only. In Fig.~\ref{fig:chm_points_B}, canopy heights are used to bias fuel placement toward areas of higher canopy. By setting $X_2$ to $-1$ at road locations and $0$ elsewhere, an effectively zero fuel probability can be given to those locations by setting $\beta_2$ large.

This section is a demonstration example of a topic requiring further development through field experience with these tools. It seems clear that a number of (potentially) available spatial covariates may be important in fuels layout, such as slope, aspect, canopy density, etc. We have seen some preliminary results that the $\beta_k$ can be identified in some circumstances, but in general increasing the problem complexity will make this difficult, likely greatly increasing the need for observation data and/or expert influence. Instead, our recommendation would be to make inferences of the heterogeneity over controlled observations, and carry those settings forward for expert tuning of the data.

%%%%%%%%%%%%%%%%%%%%%%%%%%%%%%%%%%%%%%%%%%%%%%%%%%%%%%%%%%%%%%%%%
\section{Discussion and Conclusion}

This study shows the potential of a connection from LiDAR remote sensing of mid-story to generating representative examples of the mid-story that manifest key characteristics. This is accomplished through the definition of a parameterized generative model that can capture these characteristics of the layout, and the ability to calibrate, or learn, the parameters of this model. The result is an interpretable method that can be used to compactly summarize the relevant information from collected datasets, and populate unknown areas with characteristic patterns of fuels. Although we emphasize the potential for generative model calibration, in a practical application it is reasonable to also include knowledge of the environment as either a more constrained prior parameter distribution.

A key issue is the definition of the relevant features of fuels characteristics. This work is motivated by prior investigations into environmental generative models, and reinforced by the information considered useful as embodied in historical plot survey approach to summarize the spatial heterogeneity of fuels. Further, our approach is a reasonable method for generating heterogeneous fuels layouts for 3D fire simulations by interpreting the layout of disks as locations and scaling factors for a template 3D fuel example, i.e., a shrub. The model presented is consistent with those motivations. 

The model presented should be regarded as an initial method investigated for both its flexibility and its ability to be calibrated against both perfect data and data collected in the field. There are clear extensions of this model, as well as the method of connecting the model to observations (i.e., pointcloud processing, matching metrics, and likelihood evaluation) that would be interesting, although their properties must be the subject for further investigations. Interesting extensions could include fuel type (e.g., species or coarse woody debris type) and a description of 3D properties including shape and density. In addition, further exploration of spatial covariates for fuel placement is appropriate for enhancing the ability to capture environmental conditions related to, e.g., terrain, overstory, and moisture/water. Finally, although we focus on mid-story, this generative approach can be applied to fuels from surface to canopy.  However, although the model can be extended in many ways, one of the key findings is that model complexity must be controlled to avoid models that are not feasible to infer, to specify with respect to known environment principles, and/or to interpret. 

The method of describing shape or topology, and the connection to kernels in the stochastic intensity function appears to be an open area for investigation. In application, there is limited guidance on what properties of heterogeneity and shape are relevant to environmental assessment goals, including the functional impact to wildland fire outcome. This suggests several research directions both for the underlying statistical methodology, and the connection to application pragmatic outcomes.

\section*{Acknowledgements}
\noindent\textit{This research was supported by the Los Alamos National Laboratory (LANL) through its Laboratory Directed Research and Development (LDRD) program under project number 20220024DR.}

\bibliography{FuelsGen}

\begin{thebibliography}{33}
\providecommand{\natexlab}[1]{#1}
\providecommand{\url}[1]{\texttt{#1}}
\expandafter\ifx\csname urlstyle\endcsname\relax
  \providecommand{\doi}[1]{doi: #1}\else
  \providecommand{\doi}{doi: \begingroup \urlstyle{rm}\Url}\fi

\bibitem[Anderson et~al.(2021)Anderson, Dietz, Pokswinski, Jenkins, Kaeser,
  Hiers, and Pelc]{anderson2021}
C.~Anderson, S.~Dietz, S.~Pokswinski, A.~Jenkins, M.~Kaeser, J.~Hiers, and
  B.~Pelc.
\newblock Traditional field metrics and terrestrial lidar predict plant
  richness in southern pine forests.
\newblock \emph{Forest Ecology and Management}, 491:\penalty0 119118, 2021.
\newblock ISSN 0378-1127.
\newblock \doi{https://doi.org/10.1016/j.foreco.2021.119118}.
\newblock URL
  \url{https://www.sciencedirect.com/science/article/pii/S0378112721002073}.

\bibitem[Andrews(2018)]{andrews2018}
P.~L. Andrews.
\newblock The rothermel surface fire spread model and associated developments:
  A comprehensive explanation.
\newblock \emph{Gen. Tech. Rep. RMRS-GTR-371. Fort Collins, CO: US Department
  of Agriculture, Forest Service, Rocky Mountain Research Station. 121 p.},
  371, 2018.

\bibitem[Atchley et~al.(2021)Atchley, Linn, Jonko, Hoffman, Hyman, Pimont,
  Sieg, and Middleton]{atchley2021effects}
A.~L. Atchley, R.~Linn, A.~Jonko, C.~Hoffman, J.~D. Hyman, F.~Pimont, C.~Sieg,
  and R.~S. Middleton.
\newblock Effects of fuel spatial distribution on wildland fire behaviour.
\newblock \emph{International journal of wildland fire}, 30\penalty0
  (3):\penalty0 179--189, 2021.

\bibitem[Banerjee et~al.(2003)Banerjee, Carlin, and
  Gelfand]{banerjee2003hierarchical}
S.~Banerjee, B.~P. Carlin, and A.~E. Gelfand.
\newblock \emph{Hierarchical modeling and analysis for spatial data}.
\newblock Chapman and Hall/CRC, 2003.

\bibitem[Diggle(1981)]{diggle1981binary}
P.~J. Diggle.
\newblock Binary mosaics and the spatial pattern of heather.
\newblock \emph{Biometrics}, pages 531--539, 1981.

\bibitem[Gelman et~al.(2013)Gelman, Carlin, Stern, Dunson, Vehtari, and
  Rubin]{gelmanBDA}
A.~Gelman, J.~B. Carlin, H.~S. Stern, D.~Dunson, A.~Vehtari, and D.~Rubin.
\newblock \emph{Bayesian Data Analysis}.
\newblock Chapman and Hall/CRC, 2013.

\bibitem[Hiers et~al.(2009)Hiers, O'Brien, Mitchell, Grego, and
  Loudermilk]{hiers2009}
J.~K. Hiers, J.~J. O'Brien, R.~J. Mitchell, J.~M. Grego, and E.~L. Loudermilk.
\newblock The wildland fuel cell concept: an approach to characterize
  fine-scale variation in fuels and fire in frequently burned longleaf pine
  forests.
\newblock \emph{International Journal of Wildland Fire}, 18:\penalty0 315--325,
  2009.

\bibitem[Jonko et~al.(2021)Jonko, Yedinak, Conley, and Linn]{jonko2021}
A.~K. Jonko, K.~M. Yedinak, J.~L. Conley, and R.~R. Linn.
\newblock Sensitivity of grass fires burning in marginal conditions to
  atmospheric turbulence.
\newblock \emph{Journal of Geophysical Research: Atmospheres}, 126\penalty0
  (13):\penalty0 e2020JD033384, 2021.
\newblock \doi{https://doi.org/10.1029/2020JD033384}.
\newblock URL
  \url{https://agupubs.onlinelibrary.wiley.com/doi/abs/10.1029/2020JD033384}.
\newblock e2020JD033384 2020JD033384.

\bibitem[Knapp and Keeley(2006)]{knapp2006heterogeneity}
E.~E. Knapp and J.~E. Keeley.
\newblock Heterogeneity in fire severity within early season and late season
  prescribed burns in a mixed-conifer forest.
\newblock \emph{International Journal of Wildland Fire}, 15\penalty0
  (1):\penalty0 37--45, 2006.

\bibitem[Li et~al.(2012)Li, Guo, Jakubowski, and Kelly]{li2012new}
W.~Li, Q.~Guo, M.~K. Jakubowski, and M.~Kelly.
\newblock A new method for segmenting individual trees from the lidar point
  cloud.
\newblock \emph{Photogrammetric Engineering \& Remote Sensing}, 78\penalty0
  (1):\penalty0 75--84, 2012.

\bibitem[Linn et~al.(2002)Linn, Reisner, Colman, and Winterkamp]{linn2002}
R.~Linn, J.~Reisner, J.~J. Colman, and J.~Winterkamp.
\newblock Studying wildfire behavior using firetec.
\newblock \emph{International journal of wildland fire}, 11\penalty0
  (4):\penalty0 233--246, 2002.

\bibitem[Linn et~al.(2005)Linn, Wintercamp, Colman, Edminster, and
  Bailey]{linn2005}
R.~Linn, J.~Wintercamp, J.~J. Colman, C.~Edminster, and J.~D. Bailey.
\newblock Modeling interactions between fire and atmosphere in discrete element
  fuel beds.
\newblock \emph{International journal of wildland fire}, 14\penalty0
  (1):\penalty0 37--38, 2005.

\bibitem[Linn et~al.(2012)Linn, Anderson, Winterkamp, Brooks, Wotton, Dupuy,
  Pimont, and Edminster]{linn2012}
R.~Linn, K.~Anderson, J.~Winterkamp, A.~Brooks, M.~Wotton, J.-L. Dupuy,
  F.~Pimont, and C.~Edminster.
\newblock Incorporating field wind data into firetec simulations of the
  international crown fire modeling experiment (icfme): preliminary lessons
  learned.
\newblock \emph{Canadian Journal of Forest Research}, 42\penalty0 (5):\penalty0
  879--898, 2012.

\bibitem[Linn et~al.(2020)Linn, Goodrick, Brambilla, Brown, Middleton, O'Brien,
  and Hiers]{linn2020}
R.~Linn, S.~Goodrick, S.~Brambilla, M.~Brown, R.~Middleton, J.~O'Brien, and
  J.~Hiers.
\newblock Quic-fire: A fast-running simulation tool for prescribed fire
  planning.
\newblock \emph{Environmental Modelling \& Software}, 125:\penalty0 104616,
  2020.

\bibitem[Linn et~al.(2013)Linn, Sieg, Hoffman, Winterkamp, and
  McMillin]{linn2013modeling}
R.~R. Linn, C.~H. Sieg, C.~M. Hoffman, J.~L. Winterkamp, and J.~D. McMillin.
\newblock Modeling wind fields and fire propagation following bark beetle
  outbreaks in spatially-heterogeneous pinyon-juniper woodland fuel complexes.
\newblock \emph{Agricultural and Forest Meteorology}, 173:\penalty0 139--153,
  2013.

\bibitem[Linn et~al.(2021)Linn, Winterkamp, Furman, Williams, Hiers, Jonko,
  O’Brien, Yedinak, and Goodrick]{linn2021}
R.~R. Linn, J.~L. Winterkamp, J.~H. Furman, B.~Williams, J.~K. Hiers, A.~Jonko,
  J.~J. O’Brien, K.~M. Yedinak, and S.~Goodrick.
\newblock Modeling low intensity fires: Lessons learned from 2012 rxcadre.
\newblock \emph{Atmosphere}, 12\penalty0 (2), 2021.

\bibitem[Loudermilk et~al.(2023)Loudermilk, Pokswinski, Hawley, Maxwell,
  Gallagher, Skowronski, Hudak, Hoffman, and Hiers]{lp-vegpred23}
E.~L. Loudermilk, S.~Pokswinski, C.~M. Hawley, A.~Maxwell, M.~R. Gallagher,
  N.~S. Skowronski, A.~T. Hudak, C.~Hoffman, and J.~K. Hiers.
\newblock Terrestrial laser scan metrics predict surface vegetation biomass and
  consumption in a frequently burned southeastern u.s. ecosystem.
\newblock \emph{Fire}, 6\penalty0 (4), 2023.
\newblock ISSN 2571-6255.
\newblock \doi{10.3390/fire6040151}.
\newblock URL \url{https://www.mdpi.com/2571-6255/6/4/151}.

\bibitem[McDanold et~al.(2023)McDanold, Linn, Jonko, Atchley, Goodrick, Hiers,
  Hoffman, Loudermilk, O'Brien, Parsons, Sieg, and Oliveto]{mcdanold2022}
J.~S. McDanold, R.~R. Linn, A.~K. Jonko, A.~L. Atchley, S.~L. Goodrick, J.~K.
  Hiers, C.~M. Hoffman, E.~L. Loudermilk, J.~J. O'Brien, R.~A. Parsons, C.~H.
  Sieg, and J.~A. Oliveto.
\newblock Duet - distribution of understory using elliptical transport: A
  mechanistic model of leaf litter and herbaceous spatial distribution based on
  tree canopy structure.
\newblock \emph{Ecological Modelling}, 483:\penalty0 110425, 2023.
\newblock ISSN 0304-3800.
\newblock \doi{https://doi.org/10.1016/j.ecolmodel.2023.110425}.
\newblock URL
  \url{https://www.sciencedirect.com/science/article/pii/S0304380023001564}.

\bibitem[Micheas(2019)]{cox_realizations}
A.~C. Micheas.
\newblock Cox point processes: Why one realisation is not enough.
\newblock \emph{International Statistical Review}, 87\penalty0 (2):\penalty0
  306--325, 2019.
\newblock \doi{https://doi.org/10.1111/insr.12308}.
\newblock URL \url{https://onlinelibrary.wiley.com/doi/abs/10.1111/insr.12308}.

\bibitem[Molchanov(1997)]{molchanov1997statistics}
I.~Molchanov.
\newblock \emph{Statistics of the Boolean model for practitioners and
  mathematicians}.
\newblock Wiley, 1997.

\bibitem[M{\o}ller and Helisov{\'a}(2010)]{moller2010likelihood}
J.~M{\o}ller and K.~Helisov{\'a}.
\newblock Likelihood inference for unions of interacting discs.
\newblock \emph{Scandinavian Journal of Statistics}, 37\penalty0 (3):\penalty0
  365--381, 2010.

\bibitem[M{\o}ller and Waagepetersen(2007)]{moller2007modern}
J.~M{\o}ller and R.~P. Waagepetersen.
\newblock Modern statistics for spatial point processes.
\newblock \emph{Scandinavian Journal of Statistics}, 34\penalty0 (4):\penalty0
  643--684, 2007.

\bibitem[Parsons et~al.(2017)Parsons, Linn, Pimont, Hoffman, Sauer, Winterkamp,
  Sieg, and Jolly]{parsons2017numerical}
R.~A. Parsons, R.~R. Linn, F.~Pimont, C.~Hoffman, J.~Sauer, J.~Winterkamp,
  C.~H. Sieg, and W.~M. Jolly.
\newblock Numerical investigation of aggregated fuel spatial pattern impacts on
  fire behavior.
\newblock \emph{Land}, 6\penalty0 (2):\penalty0 43, 2017.

\bibitem[Pielou(1964)]{pielou1964spatial}
E.~Pielou.
\newblock The spatial pattern of two-phase patchworks of vegetation.
\newblock \emph{Biometrics}, pages 156--167, 1964.

\bibitem[Pokswinski et~al.(2021)Pokswinski, Gallagher, Skowronski, Loudermilk,
  Hawley, Wallace, Everland, Wallace, and Hiers]{pokswinski2021}
S.~Pokswinski, M.~R. Gallagher, N.~S. Skowronski, E.~L. Loudermilk, C.~Hawley,
  D.~Wallace, A.~Everland, J.~Wallace, and J.~K. Hiers.
\newblock A simplified and affordable approach to forest monitoring using
  single terrestrial laser scans and transect sampling.
\newblock \emph{MethodsX}, 8:\penalty0 101484, 2021.
\newblock ISSN 2215-0161.
\newblock \doi{https://doi.org/10.1016/j.mex.2021.101484}.
\newblock URL
  \url{https://www.sciencedirect.com/science/article/pii/S2215016121002776}.

\bibitem[Riley et~al.(2016)Riley, Grenfell, and Finney]{riley2016}
K.~L. Riley, I.~C. Grenfell, and M.~A. Finney.
\newblock Mapping forest vegetation for the western united states using
  modified random forests imputation of fia forest plots.
\newblock \emph{Ecosphere}, 7\penalty0 (10):\penalty0 e01472, 2016.

\bibitem[Rothermel(1972)]{rothermel1972}
R.~C. Rothermel.
\newblock \emph{A mathematical model for predicting fire spread in wildland
  fuels}, volume 115.
\newblock Intermountain Forest \& Range Experiment Station, Forest Service,
  US~…, 1972.

\bibitem[Rowell et~al.(2020)Rowell, Loudermilk, Hawley, Pokswinski, Seielstad,
  Queen, O'Brien, Hudak, Goodrick, and Hiers]{rowell2020}
E.~Rowell, E.~L. Loudermilk, C.~Hawley, S.~Pokswinski, C.~Seielstad, L.~Queen,
  J.~J. O'Brien, A.~T. Hudak, S.~Goodrick, and J.~K. Hiers.
\newblock Coupling terrestrial laser scanning with 3d fuel biomass sampling for
  advancing wildland fuels characterization.
\newblock \emph{Forest Ecology and Management}, 462:\penalty0 117945, 2020.

\bibitem[Santner et~al.(2018)Santner, Williams, and Notz]{santner2018}
T.~J. Santner, B.~J. Williams, and W.~I. Notz.
\newblock \emph{The design and analysis of computer experiments, second ed.}
\newblock Springer, 2018.

\bibitem[Silva et~al.(2016)Silva, Hudak, Vierling, Loudermilk, O’Brien,
  Hiers, Jack, Gonzalez-Benecke, Lee, Falkowski, and Khosravipour]{silva2016}
C.~A. Silva, A.~T. Hudak, L.~A. Vierling, E.~L. Loudermilk, J.~J. O’Brien,
  J.~K. Hiers, S.~B. Jack, C.~Gonzalez-Benecke, H.~Lee, M.~J. Falkowski, and
  A.~Khosravipour.
\newblock Imputation of individual longleaf pine (pinus palustris mill.) tree
  attributes from field and lidar data.
\newblock \emph{Canadian Journal of Remote Sensing}, 42\penalty0 (5):\penalty0
  554--573, 2016.
\newblock \doi{10.1080/07038992.2016.1196582}.

\bibitem[Stereńczak et~al.(2020)Stereńczak, Kraszewski, Mielcarek, Żaneta
  Piasecka, Lisiewicz, and Heurich]{treesfromALS}
K.~Stereńczak, B.~Kraszewski, M.~Mielcarek, Żaneta Piasecka, M.~Lisiewicz,
  and M.~Heurich.
\newblock Mapping individual trees with airborne laser scanning data in an
  european lowland forest using a self-calibration algorithm.
\newblock \emph{International Journal of Applied Earth Observation and
  Geoinformation}, 93:\penalty0 102191, 2020.
\newblock ISSN 1569-8432.
\newblock \doi{https://doi.org/10.1016/j.jag.2020.102191}.
\newblock URL
  \url{https://www.sciencedirect.com/science/article/pii/S0303243420301665}.

\bibitem[Tinkham et~al.(2018)Tinkham, Mahoney, Hudak, Domke, Falkowski,
  Woodall, and Smith]{tinkham2018applications}
W.~T. Tinkham, P.~R. Mahoney, A.~T. Hudak, G.~M. Domke, M.~J. Falkowski, C.~W.
  Woodall, and A.~M. Smith.
\newblock Applications of the united states forest inventory and analysis
  dataset: A review and future directions.
\newblock \emph{Canadian Journal of Forest Research}, 48\penalty0
  (11):\penalty0 1251--1268, 2018.

\bibitem[Toney et~al.(2007)Toney, Rollins, Short, Frescino, Tymcio, and
  Peterson]{toney2007}
C.~Toney, M.~Rollins, K.~Short, T.~Frescino, R.~Tymcio, and B.~Peterson.
\newblock Use of fia plot data in the landfire project.
\newblock In \emph{In: McRoberts, Ronald E.; Reams, Gregory A.; Van Deusen,
  Paul C.; McWilliams, William H., eds. Proceedings of the seventh annual
  forest inventory and analysis symposium; October 3-6, 2005; Portland, ME.
  Gen. Tech. Rep. WO-77. Washington, DC: US Department of Agriculture, Forest
  Service: 309-319.}, volume~77, 2007.

\end{thebibliography}

\end{document}